%% file: main.tex
\AddToHook{package/hyperxmp/before}{%
  \RequirePackage[bookmarksnumbered,unicode]{hyperref}%
}
\documentclass[sigconf,nonacm,balance=false]{acmart}
\usepackage{amsmath}
\usepackage{booktabs}
\usepackage{tabularx}
\usepackage{graphicx}
\usepackage{placeins}
\usepackage{xcolor}
\usepackage{colortbl}
\usepackage{tikz}
\definecolor{PaperBlue}{HTML}{5271AE}
\arrayrulecolor{PaperBlue!45}
\setcopyright{none}
\acmDOI{}
\acmISBN{}
\newcommand{\bench}{AgentProofBench}
\newcommand{\pass}{\textsc{Pass}}
\newcommand{\fail}{\textsc{Fail}}
\newcommand{\unknown}{\textsc{Unknown}}
\newcommand{\error}{\textsc{Error}}
\input{revision_results.tex}



\input{strengthening_results.tex}

\title[AgentProofBench]{\texorpdfstring{AgentProofBench: Benchmarking\\Reusable Verifiers for Web Agents}{AgentProofBench: Benchmarking Reusable Verifiers for Web Agents}}

\author{Jingjie Ning}
\authornote{Corresponding author. \href{mailto:jening@cs.cmu.edu}{\texttt{jening@cs.cmu.edu}}.}
\email{jening@cs.cmu.edu}
\author{Xueqi Li}
\email{xueqil@cs.cmu.edu}
\author{Yi Lu}
\email{yilu4@cs.cmu.edu}
\author{Yibo Kong}
\email{yibok@cs.cmu.edu}
\affiliation{%
  \institution{Carnegie Mellon University}
  \department{School of Computer Science}
  \city{Pittsburgh}
  \state{Pennsylvania}
  \country{USA}}
\authorsaddresses{}
\renewcommand{\shortauthors}{Ning et al.}

\definecolor{LinkBlue}{HTML}{5271AE}
\definecolor{LinkInk}{HTML}{24292F}
\newcommand{\projecticon}{%
  \tikz[baseline=-0.6ex,x=1pt,y=1pt,line width=0.65pt,draw=LinkBlue]{%
    \draw (0,0) circle (4.6);
    \draw (-4.6,0)--(4.6,0);
    \draw (0,0) ellipse (2.2 and 4.6);
    \draw (-3.8,2.6) .. controls (-1.3,1.8) and (1.3,1.8) .. (3.8,2.6);
    \draw (-3.8,-2.6) .. controls (-1.3,-1.8) and (1.3,-1.8) .. (3.8,-2.6);}}
\newcommand{\paperlink}[3]{%
  \href{#1}{\tikz[baseline=(badge.base)]{%
    \node[rounded corners=4pt,draw=#2,fill=#2,text=white,
          inner xsep=8pt,inner ysep=4pt,font=\sffamily\small] (badge) {#3};}}}
\newcommand{\projectlink}{%
  \href{https://ethanning.github.io/AgentProofBench/}{%
    \tikz[baseline=(badge.base)]{%
      \node[rounded corners=4pt,draw=LinkBlue!30,fill=LinkBlue!7,text=LinkBlue,
            inner xsep=8pt,inner ysep=4pt,font=\sffamily\small] (badge)
       {\projecticon\hspace{5pt}Project page};}}}
\makeatletter
\def\@mkauthors{%
  \gdef\@currentauthors{}%
  \global\setbox\mktitle@bx=\vbox{%
    \unvbox\mktitle@bx\par\vskip3pt
    \centering
    {\fontsize{12}{15}\selectfont\sffamily\bfseries
      Jingjie Ning\textsuperscript{*}\quad Xueqi Li\quad Yi Lu\quad Yibo Kong\par}
    \vskip2pt
    {\normalsize School of Computer Science, Carnegie Mellon University\par}
    \vskip1pt
    {\small\texttt{\{\href{mailto:jening@cs.cmu.edu}{jening},\,
      \href{mailto:xueqil@cs.cmu.edu}{xueqil},\,
      \href{mailto:yilu4@cs.cmu.edu}{yilu4},\,
      \href{mailto:yibok@cs.cmu.edu}{yibok}\}@cs.cmu.edu}\par}
    \vskip4pt
    \projectlink\hspace{8pt}%
    \paperlink{https://github.com/EthanNing/AgentProofBench}{LinkInk}{%
      \raisebox{-2.4pt}{\includegraphics[height=11pt]{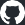}}%
      \hspace{5pt}GitHub}\par
    \vskip0pt}}
\makeatother
\hypersetup{colorlinks=true,linkcolor=LinkBlue,citecolor=LinkBlue,urlcolor=LinkBlue,
  pdftitle={AgentProofBench: Benchmarking Reusable Verifiers for Web Agents},
  pdfauthor={Jingjie Ning, Xueqi Li, Yi Lu, Yibo Kong}}

\begin{document}
\begin{abstract}
Reusable Web-agent verification requires a rule to follow changing requirements and support subsequent execution. We introduce \bench{}, which connects paired tests of a frozen acceptance rule to execution from identical saved openings. The resource spans 97 tasks, 79 families, and four Web services. Experiments with DeepSeek Flash/Pro establish two linked measurements. Across 48 families, complete-panel and decision-replay tests retain 82/96 Flash and 79/96 Pro source programs; paired reuse--rebuild BA intervals are $[-5.47,1.57]$ and $[-10.16,0.00]$ percentage points. Across 22 execution families, deployment-state panel correctness separates program assignments whose two saved openings are both delivered correctly: 32/33 assignments with error-free panels, versus 1/11 with panel errors. All reference-defined program-gate regressions and withheld deliveries occur in the latter group. Neutral continuation attains the highest correct delivery under the tested generic feedback, providing a concrete baseline for measuring the value added by a gate. Released instructions, states, programs, reference rules, and trajectories connect each test failure to its execution consequences. \bench{} makes reusable verification assessable at both the requirement and policy levels.
\end{abstract}
\begin{teaserfigure}
\centering
\includegraphics[width=\textwidth]{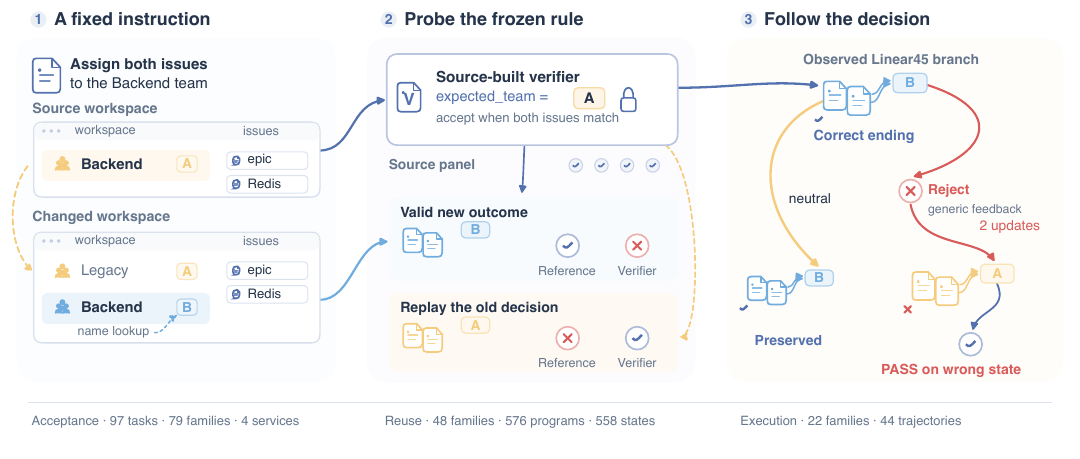}
\caption{One instruction, a changed required identity, and an acceptance rule frozen on the old identity. The illustrated Linear45 mechanism connects matched outcome tests to an observed regression after generic rejection. Team labels A/B and the displayed rule are schematic; saved code and states provide the exact instance.}
\Description{Schematic team identities A and B illustrate the Linear45 binding change. The frozen program, reference outcomes, and generic-rejection or neutral branches connect the state change to the observed execution consequence.}
\label{fig:protocol}
\end{teaserfigure}
\maketitle

\section{Introduction}
Reusable acceptance checks can make Web-agent operation fast and inspectable. A program can encode the conditions for completing a task, inspect the relevant business state, and issue the same explicit checks on subsequent executions. The central challenge is that the public instruction can stay fixed while the required work changes. The current owner of a name, the files belonging to a collection, or the value that must be preserved comes from the initial state of each execution. A verifier that remembers an earlier answer can produce a convincing decision for the wrong requirement.

The decision also becomes part of the agent's environment. Rejection can trigger further edits, and acceptance can release the resulting state. Evaluating a reusable verifier therefore requires a link between the conditions its tests exercise and the outcomes of the policy that consumes its decisions. Matched continuation histories make that link observable: the same completed work can be preserved, changed, or withheld under different stopping policies.

Existing Web-agent environments establish executable task evaluation~\cite{webarena,taubench,osworld,appworld,agentdiff}. Work on automatic assessment and generated verifiers demonstrates the importance of evidence and acceptance of valid alternatives~\cite{agentrewardbench,universal_verifier,verifierbench}. Reusable execution policies further show the value of experience across workflow instances~\cite{neurosymbolic_computer_use}. These advances motivate an evaluation that connects three properties of an acceptance rule. It must recognize the permitted terminal outcomes, recover the requirements of a changed initial state, and support effective continuation and delivery decisions.

We introduce \bench{} around this connection. An \emph{acceptance program} is executable code that judges task completion through ordinary before/after APIs. A \emph{reuse contract} fixes the public instruction and evidence permissions while initial business data vary. Matched terminal outcomes make the consequences of this variation observable. Shared actor openings then connect the same verifier to the actions taken after its decision. Every measurement is linked to versioned reference predicates, program source, API evidence, and a reconstructable business state.

The benchmark provides three contributions.
\begin{itemize}
\item A paired reuse test that holds the public instruction and generated program fixed while changing its required initial-state bindings. Complete outcome panels and replayed decisions measure complementary obligations.
\item An execution-validity test linking deployment-state panel correctness to matched policy outcomes. The 22-family study preserves 44 program assignments, two openings per assignment, and 352 policy outcomes.
\item An executable resource with 97 tasks across four Web services, including versioned reference rules, legal alternatives, reconstructable states, programs, and action histories.
\end{itemize}

The central empirical result is the connection between tests and execution. Programs with error-free deployment-state panels correctly deliver both saved openings in 32 of 33 assignments, compared with one of eleven assignments whose panels contain errors. The paired reuse study explains which requirements those panels exercise; replay witnesses expose additional errors in three source programs. Fixed-instance acceptance and cost profiles then describe the operating tradeoffs of the same verification interface. The reported generator, judge, and actor results use DeepSeek Flash/Pro with thinking disabled.

\section{Related Work}
\paragraph{Executable evaluation and generated verifiers.}
WebArena, OSWorld, $\tau$-bench, and AppWorld provide executable task environments~\cite{webarena,osworld,taubench,appworld}. Agent-Diff supplies the enterprise-service replicas used here~\cite{agentdiff}; OpenComputer and WorkForge develop verifiable software worlds and workspace criteria~\cite{opencomputer,workforge}. The test-oracle literature establishes the broader problem of trustworthy acceptance rules~\cite{testoracle}. CodeT and Eureka generate executable tests or rewards~\cite{codet,eureka}, while SpecGuardBench and WebGrader evaluate generated validation code~\cite{specguardbench,webgrader}. Rules to Tools supplies callable implementations of scientific requirements for program repair~\cite{rules_to_tools}. VerifierBench tests executable verifier generation across six agent benchmarks, including legal alternatives, ordinary errors, and attacks~\cite{verifierbench}. \bench{} extends the measurement target to matched initial-state requirements and the subsequent execution controlled by verifier decisions.

\paragraph{Evidence-based judgment.}
Agent-as-a-Judge, AJ-Bench, and AgentRewardBench study assessment from environment and trajectory evidence~\cite{agentasjudge,ajbench,agentrewardbench}. Universal Verifier develops criterion-based computer-use assessment~\cite{universal_verifier}; Sauron gathers state and trace evidence for its judgments~\cite{sauron}. VAGEN and VeriHarness study agentic verification and structured evidence collection~\cite{vagen,veriharness}. LLM-as-a-Verifier examines continuous scores, repeated evaluation, and criterion decomposition~\cite{llm_verifier}. Our fresh API judgments share the frozen states and ordinary evidence permissions of the executable programs, enabling direct comparison of valid acceptance, failure rejection, and repeated-use cost.

\begin{figure*}[t]
\centering
\includegraphics[width=\textwidth]{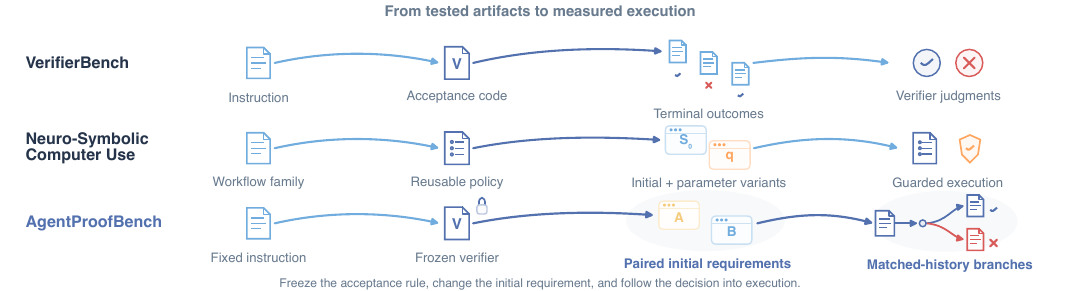}
\caption{Related settings and the measurement units in \bench{}. Paired initial requirements test a frozen acceptance program; matched histories connect its decisions to subsequent execution~\cite{verifierbench,neurosymbolic_computer_use}.}
\Description{Three illustrated workflow lanes distinguish executable verifier tests, reusable guarded execution policies, and paired verifier transfer with matched-history execution.}
\label{fig:research-scope}
\end{figure*}

\paragraph{Reuse and verification during execution.}
SkillLearnBench evaluates learned artifacts across task instances~\cite{skilllearnbench}. ControlScope studies how workflow-revision permissions shape subsequent execution~\cite{controlscope}. Revision or Re-Solving uses matched conditions to separate sources of second-pass gains~\cite{revision_resolving}. CLIFT reuses a frozen self-verification bank for training and trajectory selection~\cite{clift}. Neuro-Symbolic Computer Use learns reusable execution policies under initial-state and parameter variation, with verification-guided continuation~\cite{neurosymbolic_computer_use}. SPLIT allocates computation across sampling and imperfect verification, and verifier hardening combines adversarial improvement with protection of valid solutions~\cite{split_verification,verifier_hardening}. Enterprise-agent studies connect verification, model roles, and task completion~\cite{leni_reliability}. The Implementation Lottery compares reuse of a selected program with fresh implementations of its mechanism~\cite{implementation_lottery}. DCP compares truthful and neutral feedback from shared checkpoints in scientific-research audits~\cite{discovery_certification}. Figure~\ref{fig:research-scope} connects paired verifier tests with execution from shared histories.

\section{Benchmark Contract and Design}
\label{sec:design}
\subsection{An executable acceptance interface}
Each task supplies a public instruction $q_i$, API documentation $D_i$, and an ordinary authenticated interface $O_i$. The host reference $G_i$ evaluates task completion. For initial state $S^0_{ik}$ and terminal state $S^1_{ikj}$, a verifier produces
\begin{equation}
 \widehat y_{ikj}=V_i\!\left(q_i,O_i(S^0_{ik}),O_i(S^1_{ikj})\right),\qquad
 y_{ikj}=G_i(S^0_{ik},S^1_{ikj}).
\end{equation}
Here $k$ identifies the initial-state condition and $j$ the terminal outcome. The submitted interface is \texttt{verify(task,before,after)}. It returns \pass{}, \fail{}, or \unknown{}, with optional checks and an explanation. Exceptions, malformed output, and execution-limit exhaustion receive \error{}.

The construction prompt states the reuse contract explicitly. The program must derive state-dependent requirements from its runtime before state. Each assessment runs frozen source in a fresh read-only sandbox with the task actor's API permissions. Host predicates and terminal controls remain in the scoring harness. One execution allows one CPU, 1\,GB memory, 60 seconds, and 80 business operations across both snapshots. Expanded GraphQL root queries count separately.

\subsection{Outcome coverage and reference semantics}
Three linked studies address acceptance, reuse, and execution. The acceptance resource contains \RTasks{} tasks from \RFamilies{} families, with 20 family-disjoint development tasks. Its 556-state natural panel includes 452 successes and 104 failures, sampled by reference class with at most six states per task. The reuse study evaluates 576 programs on 558 terminal states from 48 families. The execution study compares policies on 88 saved openings from 22 families. Appendix~\ref{app:reproduction} records the natural-state selection flow and the three post-inspection task exclusions.

Four outcome panels separate common acceptance demands (Figure~\ref{fig:coverage}). Natural outcomes measure ordinary agent endings. Native-prefix pairs use service actions to connect a complete execution to an earlier incomplete state. Typed faults alter a successful anchor through service APIs, allowing the violated requirement to be identified. Legal alternatives change behavior that the public instruction permits. Successful anchors and failed states jointly test valid acceptance and failure rejection.

Reference predicates cover identities, values, collections, and relationships. Versioned selectors recover the required objects from each before state. Native-state and ordinary-API evaluation share the predicates and check restoration and evidence accessibility. Scores measure conformity to this explicit reference contract. Appendix~\ref{app:reproduction} retains the original and explicit-preservation label views.

A separate GPT-6 Astra LLM-as-a-judge audit reviews 195 states across 62 tasks, with reference verdicts withheld during initial judgment. It agrees with 191 of 192 resolved labels (99.5\%); three cases are indeterminate. The audit corroborates all 48 requirement changes, rejects all 48 failed replays, and confirms both delivered-regression pairs (Appendix~\ref{app:semantic-audit}).

The preservation task \texttt{linear\_52} illustrates the contract. Its instruction assigns Plot B3 to Rashida and requires its initial status to survive. The source state has status \texttt{Active}; the related state has \texttt{Dormant}. The reference resolves Plot B3 from the current before state, checks the requested assignee, and compares its before/after status identifiers. A correct related outcome therefore keeps \texttt{Dormant}. The instruction's other clauses create a tracker issue and update Plots A7 and C1. They supply additional matched outcomes while the B3 change isolates the state-dependent preservation requirement. The complete instruction and predicates appear in Appendix~\ref{app:reproduction}.

\subsection{Matched initial-state reuse}
The matched study, P1, selects one task from each of 48 original families, comprising 18 Linear, 10 Slack, 19 Box, and one Calendar task. Every family contributes one related initial-state transformation. Each public instruction has three initial conditions. \emph{Source} is the construction state. \emph{Background} changes data outside the requested work. \emph{Related} changes a required entity binding, initial value, collection, or preservation condition. An entity binding identifies the object denoted by a public name; a preservation condition requires an initial value to survive execution.

Qualification establishes four properties before prediction. The changed requirement follows the public instruction, service actions construct the states, native and API references agree on matched outcomes, and replaying a source-valid decision becomes invalid under the related condition. This last execution is a \emph{source-decision witness}. It tests the acceptance rule against an earlier valid business decision whose required target or value has changed.

The witness gives initial-state variation a direct semantic test. If the same earlier decision remains valid, a program could succeed by retaining its source answer. A qualified witness requires the verifier to resolve a new condition from the current before state. This separates state-dependent acceptance from incidental differences in a snapshot and makes the changed obligation inspectable in the reference and API trace.

Each initial condition contains a complete result, a distinct legal alternative where available, and two changed failures. Fault recipes and their roles remain matched across conditions; the current before state supplies identifiers and required values. Forty-two tasks have two valid outcomes per condition, and six have one. The 144 initial states support 558 main terminal states, comprising 270 valid and 288 failed entries. Witness pairs add one source-valid and one related-invalid decision per task, with their own scoring support.

DeepSeek Flash and Pro, both with thinking disabled, each generate two source programs and two rebuilt programs at each target condition, producing 576 frozen artifacts. Construction and fresh online API assessment, denoted API12, allow twelve model turns. Programs can read and debug against their construction state. Source programs run on all three conditions, while target rebuilds are compared at their own construction target. Two independent API12 judgments per model and terminal state give 2,232 online observations. The full three-construction by three-assessment matrix additionally measures reverse transfer.

\subsection{Verification during continued execution}
The execution study, P2, uses 22 families with a native completion trace containing at least two mutating operations. One Flash and one Pro actor per task generate 44 trajectories from related initial states. Every trajectory supplies a natural ending and a closed \emph{checkpoint}, saved after its first complete tool round containing a mutating request. Each of the 88 openings preserves the original initial state, current business state, and complete interaction history.

Four policies branch from each opening. \emph{Actor stop} delivers on the actor's final reply. Its non-final checkpoint outcome is a mechanical no-delivery reference. \emph{Fixed continue} allows two neutral continuation phases and delivers on a final reply after phase two. \emph{Program gate} uses a source-built Pro program, assigned by a fixed parity rule per task and actor independently of source quality. \emph{Online gate} uses fresh Pro API12 judgments. Both gates deliver on \pass{} and otherwise permit at most two continuations. The two cohorts retain the same assigned program.

The neutral message asks the actor to finish the original request while preserving completed work. Gate feedback prepends a statement of nonacceptance and carries no program checks or entity identifiers. Thus each branch evaluates a complete feedback-and-stopping policy; Appendix~\ref{app:policies} gives the exact messages. An eight-family development study selected this design before formal actor prediction. Each actor phase allows 20 model turns and 100 business operations, with a USD\,1 observed-cost stop and a 900-second deadline. Unresolved usage stops further work. Executed writes persist when delivery is withheld.

\subsection{Measurements and comparison units}
Balanced accuracy (BA) weights valid acceptance and failed-state rejection equally within each initial-state panel,
\begin{equation}
 \mathrm{BA}_{ik}=\frac12\!\left[
 \frac{\sum_{j\in J^+_{ik}}\mathbf1[\widehat y_{ikj}=\pass]}{|J^+_{ik}|}+
 \frac{\sum_{j\in J^-_{ik}}\mathbf1[\widehat y_{ikj}=\fail]}{|J^-_{ik}|}\right].
\end{equation}
$J^+_{ik}$ and $J^-_{ik}$ contain valid and failed outcomes. Repetitions are averaged within task before task averaging. \unknown{} and \error{} earn zero correctness credit. False acceptance (FA), valid acceptance, and both non-decision outcomes retain separate counts.

A \emph{source-qualified} program classifies every source-panel outcome correctly. Complete related-panel correctness tests whether this property transfers. A \emph{joint pass} additionally requires a correct witness verdict, accepting the source witness or rejecting its related replay. Main-panel correctness and joint correctness expose the number of programs that satisfy all tested conditions, alongside the average discrimination measured by BA. For the prespecified paired comparison,
\begin{equation}
 \Delta_m=\frac1N\sum_i\left(\overline{\mathrm{BA}}^{\rm reuse}_{i,r,m}
                    -\overline{\mathrm{BA}}^{\rm rebuild}_{i,r,m}\right).
\end{equation}
Unconditional results retain 96 programs per model; each conditional rate names its selected support.

P2 records correct, incorrect, withheld, and unresolved delivery for every scheduled opening. A \emph{repair} changes reference failure to success; a \emph{regression} changes success to failure. Request success, actual business change, and delivery are separate observations. The panel--delivery diagnostic groups assignments by complete related-panel correctness and records whether both openings deliver correctly. Natural endings and checkpoints are dependent views of a trajectory. Ten thousand family-bootstrap samples retain observations from the same original family; Appendix~\ref{app:statistics} specifies the primary tests.
\section{Results}
\label{sec:results}
\subsection{Joint tests of initial-state reuse}
The paired study measures whether an acceptance rule follows a changed requirement while its instruction and code remain fixed. Source programs obtain related-state BA of 93.23\% for Flash and 90.36\% for Pro (Table~\ref{tab:p1}). Reuse-minus-rebuild estimates are \SFlashDelta{} points with interval \SFlashDeltaCI{} and \SProDelta{} with \SProDeltaCI{}. Both prespecified intervals include equal mean BA; the Pro range extends to $-10.16$ points. The background control produces identical verdicts under the tested unrelated edits.

\begin{table*}[t]
\centering\small
\caption{Matched P1 results on 48 families. Values are task-averaged percentages with two program replicas or online judgments. Related reuse-minus-rebuild intervals are \SFlashDeltaCI{} for Flash and \SProDeltaCI{} for Pro.}
\label{tab:p1}
\includegraphics[width=\textwidth]{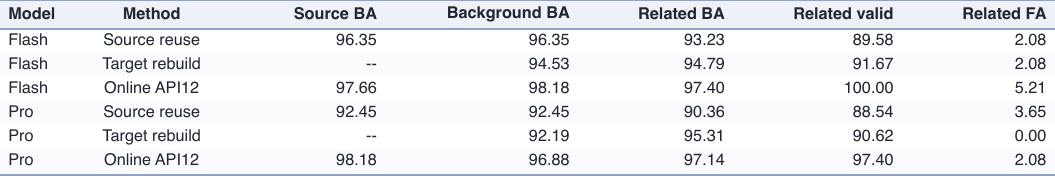}
\end{table*}

\begin{figure*}[t]
\centering
\includegraphics[width=\textwidth]{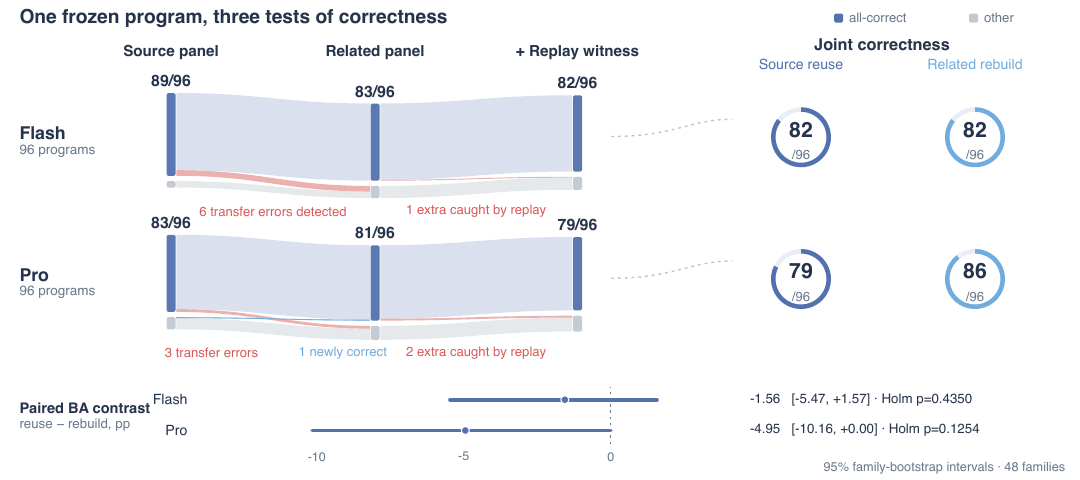}
\caption{The same 96 source programs pass through source-panel, related-panel, and joint tests. Blue denotes complete correctness and gray the remaining programs. Separate rings compare related rebuilds. The lower panel retains the paired BA contrasts and original 95\% family-bootstrap intervals.}
\Description{Related main-panel correctness is 83 of 96 Flash and 81 of 96 Pro source programs, versus 85 and 87 rebuilds. Joint correctness is 82 and 79 for source programs, versus 82 and 86 rebuilds.}
\label{fig:transfer}
\end{figure*}

Complete main-panel correctness changes from 89/96 to 83/96 source programs for Flash and from 83/96 to 81/96 for Pro. Conditional on a fully correct source panel, the retained counts are 83/89 and 80/83, spanning 45 and 46 source-qualified families. Six Flash and three Pro programs acquire an error across three families per model; one initially incorrect Pro program becomes correct on the related panel. Figure~\ref{fig:transfer} keeps the full 96-program denominator visible.

The added witness probes an earlier valid decision at its changed requirement. Source programs accept 92/96 valid source witnesses for Flash and 89/96 for Pro. After the related change, they falsely accept 6/96 and 5/96 replays, while related rebuilds accept 5/96 and 2/96. Pro source programs also give two \unknown{} and two \error{} verdicts on related witnesses. Joint correctness requires a resolved, correct verdict on both the main panel and its witness. Its resulting counts are 82/96 Flash and 79/96 Pro source programs, compared with 82/96 and 86/96 rebuilds (Figure~\ref{fig:transfer}). Rebuilt main panels are completely correct for 85/96 Flash and 87/96 Pro programs. Within the source-joint-qualified subsets, related joint correctness is 82/89 and 78/83.

\paragraph{Bindings recoverable from the current state.}
In \texttt{linear\_44}, both Flash source programs retain the old Elena identifier. Each classifies all four source outcomes correctly and rejects both valid related outcomes after the public name refers to another live user. Both related rebuilds classify the target panel correctly; one resolves the current name through the users API. The changed binding therefore has a direct witness in the before-state record, the generated source, and the returned API evidence.

In \texttt{linear\_52}, Pro replica~1 retains the source constant \texttt{Active}. It rejects correct preservation of the related state's \texttt{Dormant} value and accepts an invalid change back to \texttt{Active}. Replica~2 reads the runtime-before state and classifies all four outcomes correctly under both conditions. The difference is the acceptance relation itself, comparing the terminal status to the current initial status or to a remembered literal. The reuse contract makes that distinction testable without rewriting the user's instruction.

\paragraph{Collection dependencies beyond local faults.}
The \texttt{box\_141} instruction defines qualifying JSON through either a top-level array or a \texttt{results} array. Pro source replica~1 classifies all twelve main outcomes correctly but accepts a witness omitting a newly required JSON file. Its API read reaches the file, yet its predicate checks the wrapper's root fields and misses the record inside \texttt{results[0]}. The computed required set has eight of the nine qualifying items. Replica~2 extracts the inner record and rejects the omission.

This case explains the role of joint scoring. Passing the matched local faults establishes correctness on those outcomes. The source-decision replay then tests the acceptance of an earlier choice after a new obligation appears. Their conjunction exposes a collection branch that the main panel leaves unexercised. The published scores, omitted record, and generated predicate expose this difference for evaluating new construction methods.

\subsection{Panel correctness and execution outcomes}
\begin{figure*}[t]
\centering
\includegraphics[width=\textwidth]{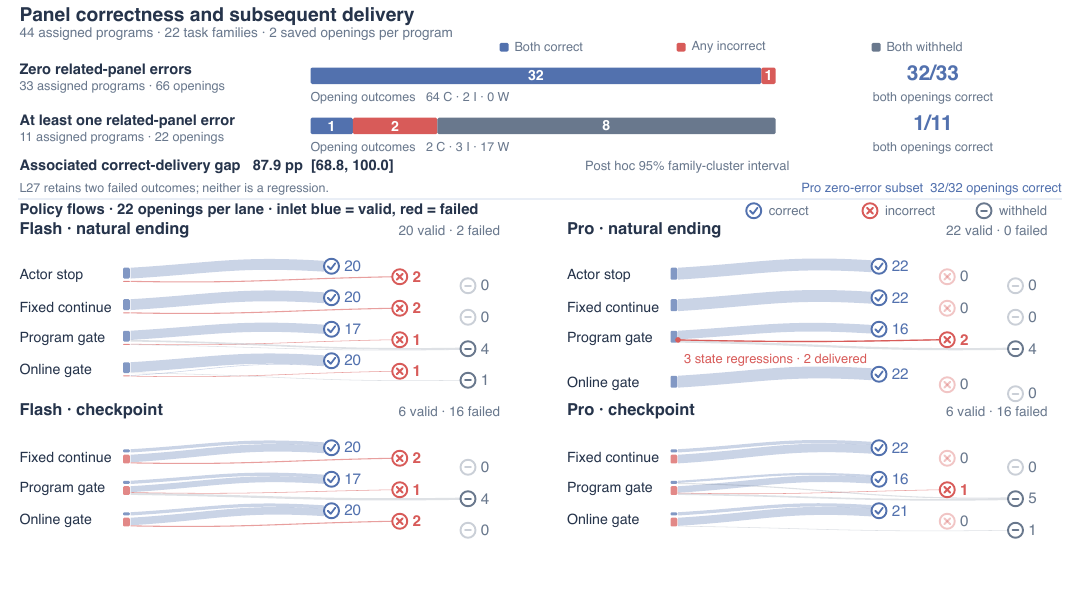}
\caption{Deployment-state panel correctness and execution outcomes. Each program assignment supplies a natural ending and checkpoint; the upper panel groups all 44 assignments by related-panel correctness. Incorrect and withheld delivery are separate from state regression. The lower policy lanes restart from 22 shared openings per actor and cohort.}
\Description{Panel-perfect programs deliver both openings correctly in 32 of 33 assignments, compared with 1 of 11 assignments containing panel errors. The policy flows retain correct, incorrect, and withheld delivery at every opening.}
\label{fig:execution}
\end{figure*}

Deployment-state tests are strongly associated with the recorded execution outcomes. Of 33 assigned programs with error-free related panels, 32 correctly deliver both their natural ending and checkpoint. Of eleven assignments with a panel error, one correctly delivers both openings (Figure~\ref{fig:execution}). At the opening level the corresponding counts are 64/66 and 2/22. The assignment is the summary unit: its two openings share one trajectory and have identical correct-versus-incorrect/withheld dispositions in this study. The 44 assignments come from 22 task families. The associated correct-delivery difference is 87.9 percentage points, with a pointwise family-cluster interval of $[68.8,100.0]$.

All seventeen program-gate withholdings and all three reference-defined regressions belong to the panel-error group. Thirteen withheld terminal states satisfy the reference, so withholding and reference failure remain distinct outcomes. The two incorrect deliveries in the error-free group are the same Flash Linear27 assignment at its two openings; both start and finish in failure. Conversely, the Flash Box130 assignment with panel errors delivers both openings correctly. Within the 36 assignments passing every source-panel test, all three with related-panel errors have incorrect or withheld delivery; 32 of the other 33 assignments deliver both openings correctly. Deployment-state labels make this a direct test of the benchmark's connection between panels and execution; operational qualification uses comparable labeled target-state panels.

\paragraph{Neutral continuation as the policy baseline.}
At checkpoints, neutral continuation repairs 14/16 Flash failures and 16/16 Pro failures, correctly delivering 20/22 and 22/22 openings. Program gating delivers 17/22 and 16/22; online gating delivers 20/22 and 21/22. Natural endings contain 20/22 Flash successes and 22/22 Pro successes. Neutral continuation preserves them without mutating requests and again delivers 20/22 and 22/22 correctly. Neutral continuation sets the highest correct-delivery level among these generic-feedback policies. Neutral continuation's paired correct-delivery gain over program gating is 13.64 points $[0.00,27.27]$ for Flash and 27.27 $[9.09,45.45]$ for Pro in each cohort.

\paragraph{Tracing an observed policy consequence (Appendix~\ref{app:policies}).}
In the Linear45 case study, a program rejects a correct assignment to the current Backend team. After generic rejection, the actor moves two issues to the old team; the program accepts and releases the resulting failure. Neutral continuation preserves the correct state. Linear41 supplies a second delivered regression after a passed source panel; Linear44 supplies a withheld case under the label-uniqueness rule. The traces expose rejection, edits, and delivery as separate events. The unit of analysis is the full gate policy, combining acceptance decisions, generic feedback, and stopping.

\subsection{Acceptance profiles on shared outcomes}
Fixed-instance tests describe the operating profile that accompanies reuse and execution. The restrictive-acceptance pattern is consistent with VerifierBench~\cite{verifierbench}: generated programs reject invalid outcomes more reliably than they accept valid alternatives. Here B4/B12 denote four-/twelve-call program construction, and API12/JSON12 denote twelve-call online judgments with ordinary/structured output. The following comparisons use Flash.

\begin{figure*}[t]
\centering
\includegraphics[width=\textwidth]{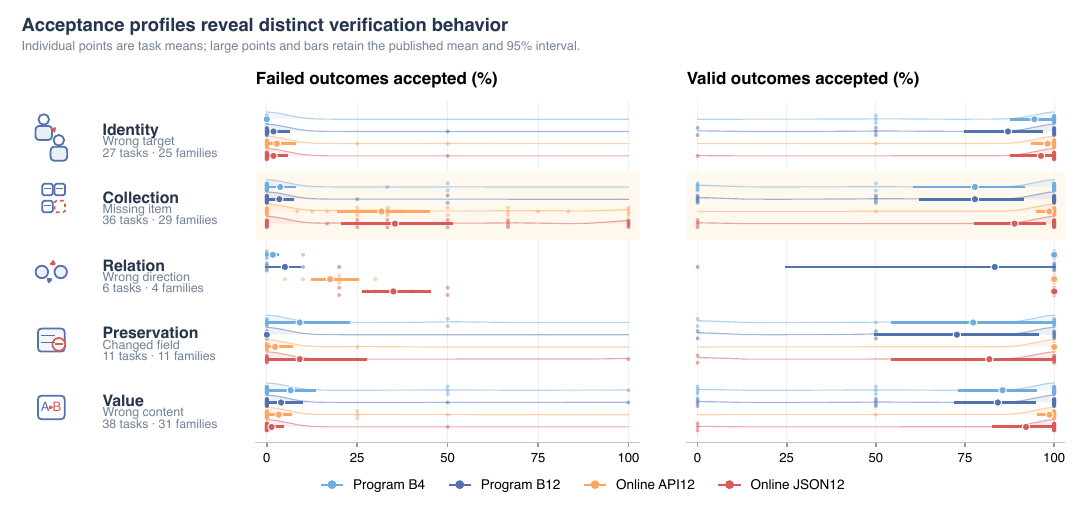}
\caption{Task-level fault-acceptance distributions. Faint curves are normalized shape guides for groups with at least ten tasks. Large points and bars retain the published means and 95\% family-bootstrap intervals.}
\Description{Program and online acceptance are compared across identity, collection completeness, relation direction, preservation, and value faults. Both failed-state rejection and valid-anchor acceptance appear for every method and category.}
\label{fig:faults}
\end{figure*}

\begin{figure*}[t]
\centering
\includegraphics[width=\textwidth]{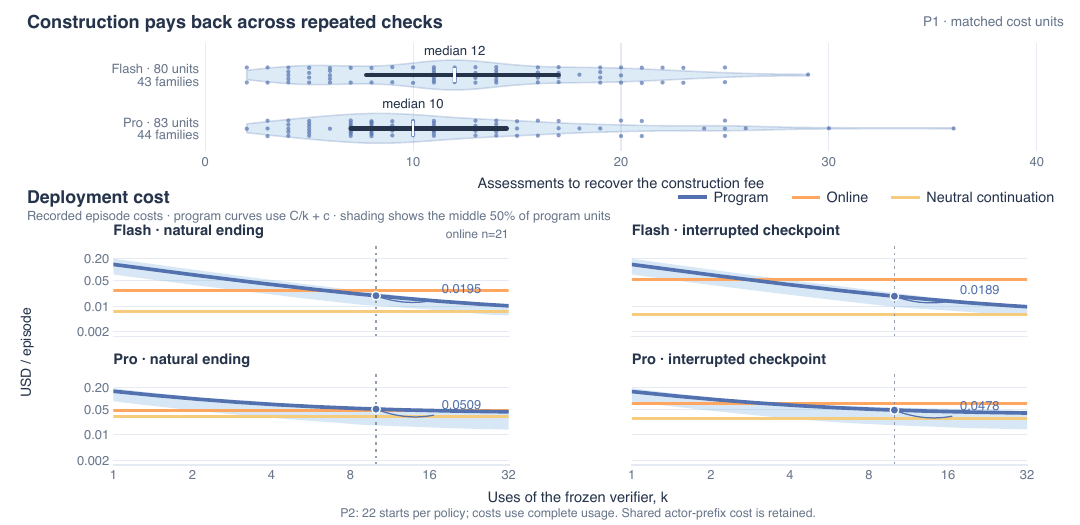}
\caption{Repeated-check costs. Top: 80/83 matched Flash/Pro units. Bottom: amortized $c+C/k$ and interquartile bands, including actor-prefix cost. Flash natural online uses 21 complete records; each other P2 cell uses 22.}
\Description{Empirical break-even distributions and four amortization curves expose construction, warm operation, and recorded comparison costs.}
\label{fig:cost}
\end{figure*}

The collection comparison uses identical before/after states and the same explicit-condition reference labels: 99 failed states and 36 successful anchors across 36 tasks and 29 families, with two observations per method. B12 false acceptance is 3.5\% versus 31.7\% for API12, a paired task-mean difference of $-28.2$ points (95\% family-bootstrap interval $[-40.7,-16.7]$). Valid-anchor acceptance is 77.8\% versus 98.6\%. Figure~\ref{fig:faults} preserves both outcomes and the distributions behind each mean.

On the shared natural panel, B12 BA is \StyleNaturalProgramCI{}\% and API12 BA is \StyleNaturalOnlineCI{}\%; the anchored hard suite gives \StyleHardProgramCI{}\% and \StyleHardOnlineCI{}\%. Median standalone assessment on the 24-state timing panel takes \StyleProgramTime{}\,s for B12 and \StyleOnlineTime{}\,s for API12. These profiles quantify selectivity, valid-solution coverage, and repeated-check latency. They complement the paired reuse and execution measurements.

\subsection{Construction cost and the reuse horizon}

Repeated use separates the fixed cost of creating a program from the cost of invoking it. The matched P1 fee comparison contains 80 Flash units from 43 families and 83 Pro units from 44 families, with known charges for all three constructions and their online judgments. Median source construction costs are USD\,0.0173/0.0765. A fresh online pass over the corresponding 9- or 12-state panel costs USD\,0.0188/0.0980. Rounding each construction-to-mean-online-fee ratio upward gives median API-fee break-even horizons of twelve assessments for Flash and ten for Pro, with ranges 2--29 and 2--36.

For P2, let $C$ be the assigned construction fee and $c$ the observed actor-prefix and branch cost. The amortized fee at $k$ uses is $c+C/k$, and warm operation costs $c$. Figure~\ref{fig:cost} shows the resulting deployment views next to online gating and neutral continuation. At ten uses, Flash program-gate means are USD\,0.0195 for natural endings and USD\,0.0189 for checkpoints. Their cold means are USD\,0.1396 and USD\,0.1390. Pro's corresponding ten-use means are USD\,0.0509 and USD\,0.0478. Construction and ongoing execution thus have distinct roles in the cost comparison.

The reuse horizon complements the acceptance and delivery profiles. A low-cost invocation is useful when its rule covers the required outcomes, and a highly discriminative judge can justify repeated inference where valid completions vary. The observed policy branches further show that construction amortization and downstream correctness are separate measurements. The released ledger counts each request once; policy views allocate the shared prefix and assigned construction explicitly. The companion retains every cold, warm, and amortized policy record.

\section{Using the Benchmark}
\bench{} supports three concrete evaluation workflows. A verifier generator can be tested on the same natural endings, legal alternatives, and typed failures to establish its acceptance profile. Its frozen outputs can then be evaluated under matched initial-state requirements using both complete-panel and joint witness correctness. Finally, the recorded actor openings support comparison of the policies that consume these decisions, measuring completed work, persistent mutations, and delivered state separately.

The resource connects these workflows through stable task, family, program, and state identities. Each false acceptance links to a violated reference condition; each transfer error links to a source program and target rebuild; each execution outcome links to the gate decision and subsequent requests. A generator improvement can therefore be assessed against the changed requirement it should recover and the policy outcome it should support.

The companion's Evidence Index provides the acceptance, reuse, and execution entry points. Canonical snapshots and exact row deltas reconstruct the saved states. The \texttt{verify(task, before, after)} interface evaluates frozen programs without model calls.

\FloatBarrier
\section{Conclusion}
\bench{} connects the two requirements of reusable Web-agent verification: recovering the current task obligation and supporting subsequent execution. Paired initial states and replay witnesses test the first; shared actor histories and program-level outcome links test the second. In the DeepSeek Flash/Pro study, deployment-state panel correctness identifies 32/33 assignments whose two openings are delivered correctly, compared with 1/11 assignments containing panel errors. Neutral continuation anchors the measurement of gate value, while acceptance profiles and amortized costs describe operating choices. The released 97-task resource makes this chain reproducible, from a changed requirement to a tested rule and its observed execution consequences.

\bibliographystyle{ACM-Reference-Format}
\bibliography{references}
\appendix

\begin{figure*}[t]
\centering
\includegraphics[width=\textwidth]{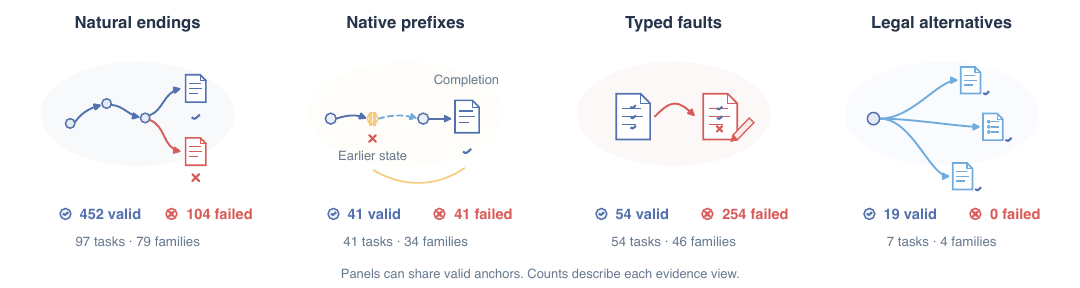}
\caption{Four complementary evidence views. Counts refer to each panel; successful anchors can be shared. The natural panel uses the frozen six-state cap and has 72 dual-class tasks.}
\Description{Illustrated natural trajectories, native prefixes, targeted faults, and legal alternatives show valid/failed counts and task/family support.}
\label{fig:coverage}
\end{figure*}

\section{Reproduction and References}
\label{app:reproduction}
The environment uses Agent-Diff revision \texttt{3bb9c407}. Snapshot clients expose \texttt{request}, \texttt{get}, \texttt{post}, and \texttt{graphql}; the host supplies authentication and mounts the public catalog at \path{/input/api_docs.json}. Python 3.12 sandboxes provide a read-only filesystem, bounded scratch space, and an unprivileged user. Qualification can act through the original content author's ordinary APIs, then restores the task actor before saving the state.

The companion's Evidence Index maps the acceptance corpus, paired reuse panels, execution openings, cross-state judgments, and their frozen protocols to canonical artifacts. Its README provides the assessment and reproduction commands. Original code and annotations use MIT; upstream notices are included.

The natural-state selection flow is 776 resolved attempts $\rightarrow$ 697 unique terminal snapshots $\rightarrow$ 556 main-panel states (452 successes, 104 failures). The 141 remaining unique states contain 138 successes and three failures and enter the all-attempt sensitivity. The main panel samples at most six states per task by reference class, independently of verifier predictions.

The explicit-acceptance view excludes \texttt{slack\_105}, \texttt{box\_159}, and \texttt{calendar\_172} after inspection of earlier results. Their undeclared boundaries concern a completion-text span, formatting around preserved sections, and an event title versus attendee phrase. All original 100-task outputs remain available. The explicit-preservation view changes 19 typed-state labels across seven Linear tasks, reducing negative-task support from 61 to 54.

Program construction and repeated online judgment have separately frozen execution protocols. Their reported collection comparison uses the same 99 failed states, 36 successful anchors, before/after references, and explicit-condition labels across 36 tasks and 29 families. The shared label map combines the eligible base manifest with the typed extension; both methods have two observations per state. The extension increases total typed negatives from 177 to 254 across categories. Both execution protocols are scored against this shared input set. The paired collection contrasts are descriptive family-bootstrap intervals; the prespecified hypothesis tests concern reuse--rebuild BA.

\paragraph{LLM-as-a-judge semantic audit.}
\label{app:semantic-audit}
GPT-6 Astra (\texttt{gpt-6-astra}, \texttt{xhigh}) reviewed a pre-existing stratified packet in three parallel fresh-context sessions, with one initial judgment per case. Inputs comprised the public instruction, before/after business tables, and comparison initial state when available. Python supported retrieval, identity resolution, byte decoding, and arithmetic. Reference labels and verifier outputs were withheld until all judgments were saved. The 195 unique states cover 62 tasks: 48 related completions, 48 replays, 66 source/reference disagreements, twelve program/online disagreements, six regression endpoints, and four states from each outcome panel; one state belongs to two strata.

Agreement is 191/192 among resolved cases (99.5\%, Cohen's $\kappa=0.989$), or 191/195 over the entire packet. Three cases are indeterminate: two involve quotation delimiters, and one requires Range-read process evidence beyond its satisfied terminal condition. The resolved disagreement is Linear44's withheld ending: the reference requires one live \texttt{hardware-failure} label, while Astra accepts the requested issue with an additional unused matching label. Its delivery remains withheld under either interpretation. Both delivered regressions, Linear41 and Linear45, agree with the semantic review. The released audit preserves every first-pass label, reason, reference comparison, and reconciliation; benchmark scores retain the frozen reference labels.

\paragraph{A complete instruction-to-reference example.}
The original \texttt{linear\_52} instruction is reproduced below.
\begin{quote}\small
Create a new Garden Plots issue titled `Spring 2025 Plot Reassignment Tracker' with description `Documenting all plot changes for the growing season. Reassignments finalized at March board meeting.'. Set the existing `Plot A7 - Tomatoes' assignee to Ines and status to Active. Set the existing `Plot B3 - Herbs' assignee to Rashida and keep its initial status. Set the existing `Plot C1 - Squash' status to Dormant and leave it unassigned.
\end{quote}
The reference requires the new issue and exact description, the requested A7 assignee and status, the B3 assignee and preserved initial status, and the C1 status and empty assignee. The related-state preparation changes B3 from \texttt{Active} to \texttt{Dormant} before execution. Its preservation predicate becomes $\mathrm{status}_{after}(B3)=\mathrm{status}_{before}(B3)=\texttt{Dormant}$. Returning B3 to \texttt{Active} fails this predicate. The complete before/after states and API observations are linked in the released case record.

Dynamic qualification searches 73 candidate tasks from 64 original families, retains 57 tasks from 51 families, and selects the formal 48-family cohort using the frozen inventory order and service coverage rule. Distinctness requires different business behavior. Duplicate legal outcomes are omitted consistently across all three initial states; the six single-positive tasks retain one valid and two failed roles in every condition. Source tag sets are empty for Box123 and Box163, so their replay witnesses test retention of newly present tags. Box117's main fault changes a folder name and its witness tests the former parent binding. Box155's panel has one CSV per domain; the full record identifies the corresponding branch coverage.

\section{Statistical Protocol}
\label{app:statistics}
The family bootstrap uses 10,000 resamples and averages replicas within task. Complete common panels define each paired estimate. \unknown{} and \error{} remain in its denominator with zero correctness credit. BA requires valid and failed outcomes; source-witness acceptance and related-witness false acceptance use their respective classes. The two prespecified related reuse-minus-rebuild contrasts use centered-bootstrap tests with Holm adjustment. Controlled panels measure discrimination under their declared outcome mix; natural cohorts measure the recorded actor endings. Figure~\ref{fig:transfer} shows their intervals; Holm-adjusted $p$ values are 0.4350 for Flash and 0.1254 for Pro. Source changes, strata, witness counts, cross-construction results, and execution-policy comparisons are additional diagnostics with pointwise intervals. The panel--delivery association is an exploratory analysis of frozen assignments. Its bootstrap resamples 22 task families, retaining both actor assignments and both openings; 9,999 of 10,000 draws contain both panel groups. The statistic is the difference in the fraction of assignments whose two openings are both delivered correctly.

The program-level transition table uses all 96 source programs per model. Flash has 83 correct on both source and related panels, six correct on source only, and seven with errors on both. Pro has 80 correct on both, three correct on source only, one correct on related only, and twelve with errors on both. Background changes preserve all source-qualified programs. Unconditional counts retain the syntax-failing Pro program from Box130.

\begin{figure*}[t]
\centering
\includegraphics[width=\textwidth]{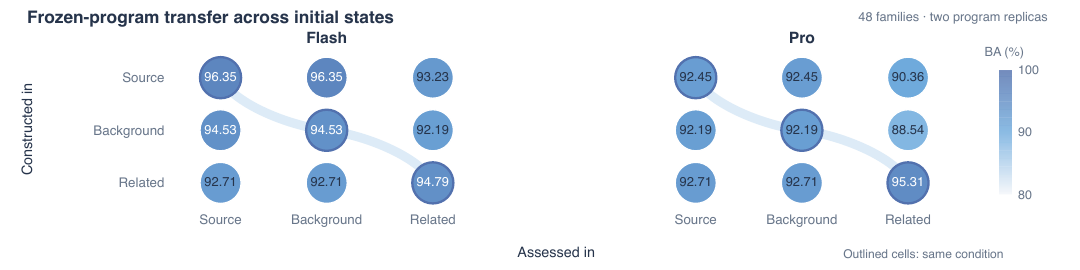}
\caption{The complete construction/assessment matrix. Circle color gives task-averaged BA on a common 80--100\% scale; all 18 values retain two replicas and 48 families. Outlined cells assess the same condition used for construction. The matrix contains 6,696 program predictions.}
\Description{Two three-by-three circle maps show directional cross-condition performance for Flash and Pro.}
\label{fig:cross}
\end{figure*}

Figure~\ref{fig:cross} measures forward and reverse transfer for every construction condition. The Evidence Index links the six-category breakdown and family supports. The released tables also include valid acceptance, false acceptance, \unknown{}, \error{}, program agreement, and witness verdict pairs. Source-witness conditional acceptance has 92 Flash and 89 Pro qualifying programs, spanning 47 and 48 families. These differ from the source-panel supports of 89 and 83 programs, spanning 45 and 46 families, because the two diagnostics establish different prerequisites. Related rebuilds accept 5/96 Flash and 2/96 Pro failed replay witnesses.

\section{Policy Outcomes and Costs}
\label{app:policies}
The eight-family development gate uses sixteen trajectories and one continuation under neutral, generic-checklist, or public-evidence feedback. Integrity must hold on at least six families, with successful and failed openings each covering at least three families. A feedback arm qualifies through changes on at least six families and reference transitions on at least five distinct trajectories. Neutral feedback meets these criteria with changes on seven families and repairs on thirteen trajectories, selecting the formal four-policy design. The formal checkpoint rule retains failed, idempotent, and already-complete mutation rounds, and all 44 trajectories supply an available checkpoint.

Figure~\ref{fig:execution} gives the 352 policy outcomes, with separate natural and checkpoint cohorts; Figure~\ref{fig:cost} gives the cost views. Cold, warm, and ten-use fees retain the shared actor prefix and use identical complete-usage records. One Linear29 connection timeout leaves the Flash natural online fee unknown and withholds a successful outcome.

\noindent\begin{minipage}{\linewidth}
Each fixed-continuation phase receives this message.
\begin{quote}\small
Continue checking and completing the original user request. Preserve valid work already completed.
\end{quote}
The two gate policies receive the following message after nonacceptance.
\begin{quote}\small
The acceptance check has not accepted this outcome. Continue checking and completing the original user request. Preserve valid work already completed.
\end{quote}
\end{minipage}

Source-panel qualification retains 17 Flash and 19 Pro assignments. Program gating and neutral continuation each correctly deliver 16/17 Flash assignments in each cohort; all four Flash withholdings occur outside this source-qualified group. For Pro the corresponding counts are 16/19 and 19/19. Two natural regressions remain after source qualification. Deployment-state panel correctness supplies the finer separation reported in the main text.

On a failed Linear41 Flash checkpoint, the online gate returns \pass{} and fresh Pro returns \fail{} for identical snapshots.

These messages use the same original task and saved interaction history. The source-qualified analysis filters existing assignments and preserves their observed outcomes. The related-rebuilt verdicts use existing same-state cross-checks; they measure an alternative initial acceptance decision.

There are 278 actual continuation phases. At natural endings, program gating sends 19 Flash and nine Pro mutating requests, with fifteen and nine successes. At checkpoints, fixed continuation sends 49/57 requests with 45/57 successes for Flash/Pro; program gating sends 49/65 with 48/65 successes, and online gating sends 47/62 with 45/62 successes. The three Pro natural regressions occur in Linear41, Linear44, and Linear45, with four, three, and two successful mutations. The case index links all three histories, source checks, and reference failures.

The cross diagnostic records 2,484 program scores and 195 fresh Pro judgments over 207 states. Assigned source-Pro versus fresh-Pro agreement is 81.82\%/75.00\% at natural endings and 85.23\%/76.14\% at checkpoints for Flash/Pro actors, after averaging within anchor and task. The online-gate comparison uses 124 matched observations on 121 decoded state pairs from 87 anchors. Its two unmatched intermediate observations belong to the program-only set. The resource retains their identities and all four-by-four verdict counts.

Both experimental models use temperature zero and thinking disabled, requesting \texttt{deepseek-flash} and \texttt{deepseek-v4-pro}. Construction and API12 allow 32,768 output tokens per turn; actors allow 4,096 and development public-feedback investigation allows 2,048. The ledger records requested and returned aliases and UTC times on October 24--25, 2026. Reference development used \texttt{gpt-6.1-sol} with \texttt{xhigh} reasoning effort.

The released ledger retains retries, unresolved usage, interrupted construction, and reconstruction provenance. These records support reproduction and charge reconciliation separately from the policy cost comparisons.

\begin{figure*}[t]
\centering
\includegraphics[width=\textwidth]{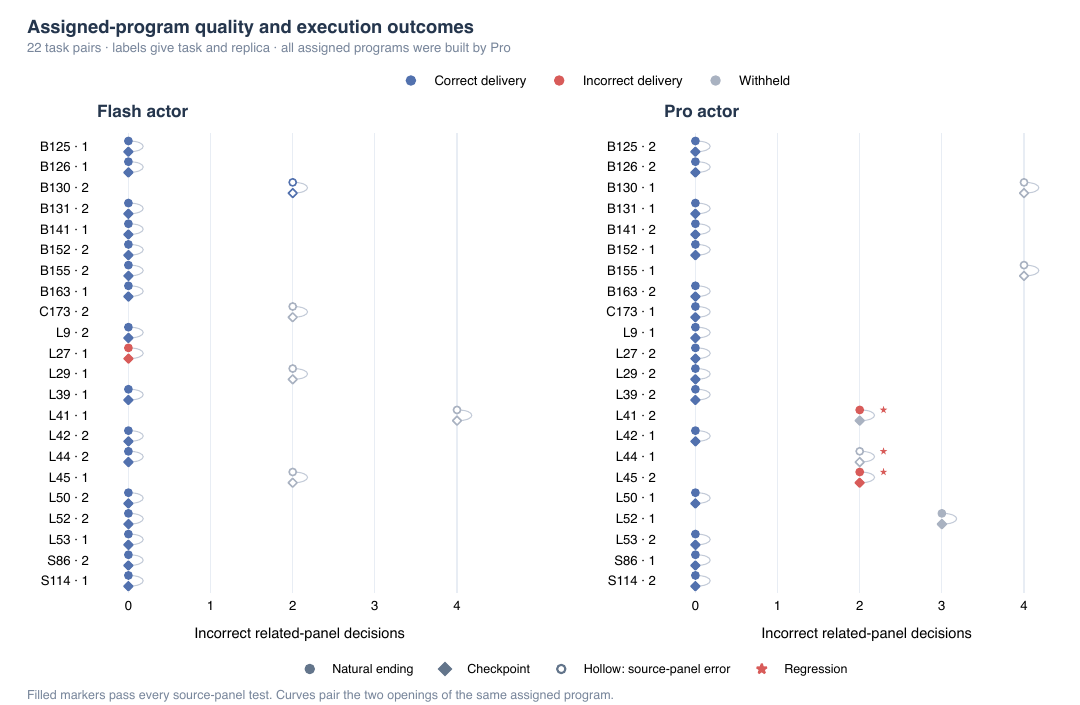}
\caption{Every assigned program linked to its execution outcomes. Horizontal position gives related-panel errors; filled markers pass every source-panel test. Circles show natural endings, diamonds checkpoints, and stars mark reference-defined regressions. Color gives correct, incorrect, or withheld delivery. Labels identify task and replica; curves pair openings of the same assignment.}
\Description{All 22 task pairs retain source qualification, related-panel errors, replica identity, and both cohorts of observed delivery.}
\label{fig:program-landscape}
\end{figure*}

\section{Construction and Archives}
\label{app:archive}
In the original construction study, SelfReview uses the instruction and documentation, InitialRead adds initial-state API reads, and Bounded adds execution feedback. B4 and B12 denote four- and twelve-call construction budgets. InitialRead minus SelfReview improves natural BA by 12.0 [4.2, 20.4] points for Flash and 12.3 [4.2, 20.8] for Pro. The resource retains the thinking-mode, construction-budget, and workflow comparisons with their full intervals.

The source diagnosis inspects 149 false rejections from 22 programs: 109 representation/identifier cases, eight omitted legal alternatives, seven exact-match constraints, nine unrequested preservation conditions, and sixteen other task-binding errors. Rule-specific edits assisted by \texttt{gpt-6.1-sol} change all 149 inspected decisions to acceptance. The eighteen available negative controls return \fail{} afterward, compared with seventeen beforehand; four affected programs have no eligible control. This diagnostic evaluates the edits on the inspected cases and available negative controls; replay uses the saved artifacts. All benchmark scores use the original frozen predictions.

The archived pilot contains twelve tasks from eleven families, a matched four-state target panel, a separate two-state source probe, and 96 policy outcomes. Its 34 continuations make zero mutating requests. The formal experiment uses its own family cohort, matched three-condition panels, and saved checkpoint branches.
\end{document}

%% file: revision_results.tex
\newcommand{\RTasks}{97}
\newcommand{\RFamilies}{79}

\newcommand{\StyleNaturalProgramCI}{92.1 [87.7, 95.9]}
\newcommand{\StyleNaturalOnlineCI}{99.0 [97.7, 99.9]}
\newcommand{\StyleHardProgramCI}{87.1 [80.6, 93.6]}
\newcommand{\StyleHardOnlineCI}{96.2 [93.8, 98.1]}

\newcommand{\StyleProgramTime}{0.58}
\newcommand{\StyleOnlineTime}{20.69}

%% file: strengthening_results.tex
\newcommand{\SFlashDelta}{-1.56}
\newcommand{\SFlashDeltaCI}{[-5.47, 1.57]}

\newcommand{\SProDelta}{-4.95}
\newcommand{\SProDeltaCI}{[-10.16, 0.00]}

%% file: references.bib
@misc{implementation_lottery,
  title = {{The Implementation Lottery: Auditing Idea Reliability in Automated Research}},
  author = {Ning, Jingjie and Zhong, Shanshan and Li, Xiaochuan and Zeng, Ji and Xiong, Chenyan},
  year = {2026},
  eprint = {2607.26587},
  archivePrefix = {arXiv},
  doi = {10.48550/arXiv.2607.26587},
  url = {https://arxiv.org/abs/2607.26587}
}

@misc{discovery_certification,
  title = {{Scores Alone Do Not Prove Discovery: The Discovery Certification Protocol for Auditing AI Research Agents}},
  author = {Ning, Jingjie and Zhong, Shanshan and Li, Xiaochuan and Zeng, Ji},
  year = {2026},
  eprint = {2609.09219},
  archivePrefix = {arXiv},
  doi = {10.48550/arXiv.2609.09219},
  url = {https://arxiv.org/abs/2609.09219}
}

@misc{rules_to_tools,
  title = {{Rules to Tools: Executable Checks for LLM Agents in Scientific Computing}},
  author = {Ning, Jingjie and Zhao, Guojiang and Xu, Chen and Zhong, Shanshan and Li, Xiaochuan and Zeng, Ji and Ke, Guolin},
  year = {2026},
  eprint = {2610.00313},
  archivePrefix = {arXiv},
  doi = {10.48550/arXiv.2610.00313},
  url = {https://arxiv.org/abs/2610.00313}
}

@misc{controlscope,
  title = {{ControlScope: Workflow Revision and Reliability in LLM Agents}},
  author = {Ning, Jingjie and Li, Xueqi and Kong, Yibo and Li, Dongting},
  year = {2026},
  eprint = {2609.34313},
  archivePrefix = {arXiv},
  doi = {10.48550/arXiv.2609.34313},
  url = {https://arxiv.org/abs/2609.34313}
}

@misc{revision_resolving,
  title = {{Revision or Re-Solving? Decomposing Second-Pass Gains in Multi-LLM Pipelines}},
  author = {Ning, Jingjie and Li, Xueqi and Yu, Chengyu},
  howpublished = {Conference on Language Modeling (COLM)},
  year = {2026},
  eprint = {2604.01029},
  archivePrefix = {arXiv},
  doi = {10.48550/arXiv.2604.01029},
  url = {https://arxiv.org/abs/2604.01029}
}

@misc{webarena,
  title = {{WebArena: A Realistic Web Environment for Building Autonomous Agents}},
  author = {Zhou, Shuyan and Xu, Frank F. and Zhu, Hao and Zhou, Xuhui and Lo, Robert and Sridhar, Abishek and Cheng, Xianyi and Ou, Tianyue and Bisk, Yonatan and Fried, Daniel and Alon, Uri and Neubig, Graham},
  year = {2023},
  eprint = {2307.13854},
  archivePrefix = {arXiv},
  doi = {10.48550/arXiv.2307.13854},
  url = {https://arxiv.org/abs/2307.13854}
}

@misc{taubench,
  title = {{$\tau$-bench: A Benchmark for Tool-Agent-User Interaction in Real-World Domains}},
  author = {Yao, Shunyu and Shinn, Noah and Razavi, Pedram and Narasimhan, Karthik},
  year = {2024},
  eprint = {2406.12045},
  archivePrefix = {arXiv},
  doi = {10.48550/arXiv.2406.12045},
  url = {https://arxiv.org/abs/2406.12045}
}

@misc{osworld,
  title = {{OSWorld: Benchmarking Multimodal Agents for Open-Ended Tasks in Real Computer Environments}},
  author = {Xie, Tianbao and Zhang, Danyang and Chen, Jixuan and Li, Xiaochuan and Zhao, Siheng and Cao, Ruisheng and Hua, Toh Jing and Cheng, Zhoujun and Shin, Dongchan and Lei, Fangyu and Liu, Yitao and Xu, Yiheng and Zhou, Shuyan and Savarese, Silvio and Xiong, Caiming and Zhong, Victor and Yu, Tao},
  year = {2024},
  eprint = {2404.07972},
  archivePrefix = {arXiv},
  doi = {10.48550/arXiv.2404.07972},
  url = {https://arxiv.org/abs/2404.07972}
}

@misc{agentasjudge,
  title = {{Agent-as-a-Judge: Evaluate Agents with Agents}},
  author = {Zhuge, Mingchen and Zhao, Changsheng and Ashley, Dylan and Wang, Wenyi and Khizbullin, Dmitrii and Xiong, Yunyang and Liu, Zechun and Chang, Ernie and Krishnamoorthi, Raghuraman and Tian, Yuandong and Shi, Yangyang and Chandra, Vikas and Schmidhuber, J{\"u}rgen},
  year = {2024},
  eprint = {2410.10934},
  archivePrefix = {arXiv},
  doi = {10.48550/arXiv.2410.10934},
  url = {https://arxiv.org/abs/2410.10934}
}

@misc{eureka,
  title = {{Eureka: Human-Level Reward Design via Coding Large Language Models}},
  author = {Ma, Yecheng Jason and Liang, William and Wang, Guanzhi and Huang, De-An and Bastani, Osbert and Jayaraman, Dinesh and Zhu, Yuke and Fan, Linxi and Anandkumar, Anima},
  year = {2023},
  eprint = {2310.12931},
  archivePrefix = {arXiv},
  doi = {10.48550/arXiv.2310.12931},
  url = {https://arxiv.org/abs/2310.12931}
}

@misc{codet,
  title = {{CodeT: Code Generation with Generated Tests}},
  author = {Chen, Bei and Zhang, Fengji and Nguyen, Anh and Zan, Daoguang and Lin, Zeqi and Lou, Jian-Guang and Chen, Weizhu},
  year = {2022},
  eprint = {2207.10397},
  archivePrefix = {arXiv},
  doi = {10.48550/arXiv.2207.10397},
  url = {https://arxiv.org/abs/2207.10397}
}

@misc{specguardbench,
  title = {{SpecGuardBench: Can Coding Agents Construct Reliable Validation Harnesses from Natural Language Specifications?}},
  author = {Zhong, Zhiqing and Chen, Mingyu and Fang, Aoyang and Zhang, Yuzhong and Tian, Haoye and Briand, Lionel and Yu, Boxi and He, Pinjia},
  year = {2026},
  eprint = {2610.16870},
  archivePrefix = {arXiv},
  doi = {10.48550/arXiv.2610.16870},
  url = {https://arxiv.org/abs/2610.16870}
}

@misc{ajbench,
  title = {{AJ-Bench: Benchmarking Agent-as-a-Judge for Environment-Aware Evaluation}},
  author = {Shi, Wentao and Wang, Yu and Zhao, Yuyang and Chen, Yuxin and Feng, Fuli and Hao, Xueyuan and Su, Xi and Gu, Qi and Su, Hui and Cai, Xunliang and He, Xiangnan},
  year = {2026},
  eprint = {2604.18240},
  archivePrefix = {arXiv},
  doi = {10.48550/arXiv.2604.18240},
  url = {https://arxiv.org/abs/2604.18240}
}

@misc{webgrader,
  title = {{WebGrader: Training LLMs for Web Development with Self-Evolving Programmatic Grader}},
  author = {Chen, Boshui and Liu, Huiping and Zhang, Shaolei},
  year = {2026},
  eprint = {2608.06474},
  archivePrefix = {arXiv},
  doi = {10.48550/arXiv.2608.06474},
  url = {https://arxiv.org/abs/2608.06474}
}

@misc{vagen,
  title = {{Agentic Reward Modeling: Verifying GUI Agent via Progressive Trajectory-Grounded Interaction}},
  author = {Cui, Chaoqun and Huang, Jing and Wang, Shijing and Zheng, Liming and Kong, Qingchao and Zeng, Zhixiong},
  year = {2026},
  eprint = {2602.00575},
  archivePrefix = {arXiv},
  doi = {10.48550/arXiv.2602.00575},
  url = {https://arxiv.org/abs/2602.00575}
}

@misc{agentrewardbench,
  title = {{AgentRewardBench: Evaluating Automatic Evaluations of Web Agent Trajectories}},
  author = {L\`u, Xing Han and Kazemnejad, Amirhossein and Meade, Nicholas and Patel, Arkil and Shin, Dongchan and Zambrano, Alejandra and Sta\'nczak, Karolina and Shaw, Peter and Pal, Christopher J. and Reddy, Siva},
  year = {2025},
  eprint = {2504.08942},
  archivePrefix = {arXiv},
  doi = {10.48550/arXiv.2504.08942},
  url = {https://arxiv.org/abs/2504.08942}
}

@misc{appworld,
  title = {{AppWorld: A Controllable World of Apps and People for Benchmarking Interactive Coding Agents}},
  author = {Trivedi, Harsh and Khot, Tushar and Hartmann, Mareike and Manku, Ruskin and Dong, Vinty and Li, Edward and Gupta, Shashank and Sabharwal, Ashish and Balasubramanian, Niranjan},
  year = {2024},
  eprint = {2407.18901},
  archivePrefix = {arXiv},
  doi = {10.48550/arXiv.2407.18901},
  url = {https://arxiv.org/abs/2407.18901}
}

@misc{skilllearnbench,
  title = {{SkillLearnBench: Benchmarking Continual Learning Methods for Agent Skill Generation on Real-World Tasks}},
  author = {Zhong, Shanshan and Lu, Yi and Ning, Jingjie and Wan, Yibing and Feng, Lihan and Ao, Yuyi and Ribeiro, Leonardo F. R. and Dreyer, Markus and Ammirati, Sean and Xiong, Chenyan},
  year = {2026},
  eprint = {2604.20087},
  archivePrefix = {arXiv},
  doi = {10.48550/arXiv.2604.20087},
  url = {https://arxiv.org/abs/2604.20087}
}

@inproceedings{agentdiff,
  author = {Pysk{\l}o, Hubert M. and Zhuravel, Artem and Watson, Patrick D.},
  title = {{Agent-Diff: Benchmarking LLM Agents on Enterprise API Tasks via Code Execution with State-Diff-Based Evaluation}},
  booktitle = {Proceedings of the 32nd ACM SIGKDD Conference on Knowledge Discovery and Data Mining},
  year = {2026},
  publisher = {Association for Computing Machinery},
  address = {New York, NY, USA},
  doi = {10.1145/3770855.3817555},
  url = {https://doi.org/10.1145/3770855.3817555},
  pages = {9707--9717},
  numpages = {11}
}

@misc{universal_verifier,
  author = {Rosset, Corby and Sharma, Pratyusha and Zhao, Andrew and Gonzalez-Fernandez, Miguel and Awadallah, Ahmed},
  title = {{The Art of Building Verifiers for Computer Use Agents}},
  year = {2026},
  eprint = {2604.06240},
  archivePrefix = {arXiv},
  doi = {10.48550/arXiv.2604.06240},
  url = {https://arxiv.org/abs/2604.06240}
}

@misc{veriharness,
  title = {{VeriHarness: Scaling Agentic Verification for Long-Horizon Tasks}},
  author = {Zhang, Caiqi and Han, Rujun and Wang, Zifeng and CuiZhu, Zoey and Collier, Nigel and Pfister, Tomas and Lee, Chen-Yu},
  year = {2026},
  eprint = {2610.00972},
  archivePrefix = {arXiv},
  doi = {10.48550/arXiv.2610.00972},
  url = {https://arxiv.org/abs/2610.00972}
}

@misc{opencomputer,
  title = {{OpenComputer: Verifiable Software Worlds for Computer-Use Agents}},
  author = {Wei, Jinbiao and Ma, Qianran and Zhao, Yilun and Zhou, Xiao and Ni, Kangqi and Gan, Guo and Cohan, Arman},
  year = {2026},
  eprint = {2605.19769},
  archivePrefix = {arXiv},
  doi = {10.48550/arXiv.2605.19769},
  url = {https://arxiv.org/abs/2605.19769}
}

@misc{workforge,
  title = {{Scaling Verifiable Environments for Long-horizon Work Agents}},
  author = {Zhang, Jiazheng and Ma, Long and Yang, Yunxian and Xi, Zhiheng and Lei, Zhikai and Yang, Yajie and Liao, Chenyang and Zhou, Enyu and Nan, Yang and Tian, Yuchen and Jin, Senjie and Wang, Yibo and He, Wei and Liu, Boyang and Huang, Jixuan and Guo, Xin and Jiang, Shutong and Hao, Zhezheng and Liang, Xinbing and Zhang, Zhihao and Zhou, Changzhi and Zhou, Wiggin and Gui, Tao and Zhang, Qi and Huang, Xuanjing and Clarenceai},
  year = {2026},
  eprint = {2610.04906},
  archivePrefix = {arXiv},
  doi = {10.48550/arXiv.2610.04906},
  url = {https://arxiv.org/abs/2610.04906}
}

@article{testoracle,
  title = {{The Oracle Problem in Software Testing: A Survey}},
  author = {Barr, Earl T. and Harman, Mark and McMinn, Phil and Shahbaz, Muzammil and Yoo, Shin},
  journal = {IEEE Transactions on Software Engineering},
  year = {2015},
  volume = {41},
  number = {5},
  pages = {507--525},
  doi = {10.1109/TSE.2014.2372785},
  url = {https://doi.org/10.1109/TSE.2014.2372785}
}

@misc{split_verification,
  title = {{Sample, Mix, or Verify? Allocating Test-Time Compute under Imperfect Verifiers}},
  author = {Iyengar, Srinivasan},
  year = {2026},
  eprint = {2610.23255},
  archivePrefix = {arXiv},
  doi = {10.48550/arXiv.2610.23255},
  url = {https://arxiv.org/abs/2610.23255}
}

@misc{neurosymbolic_computer_use,
  title = {{Neuro-Symbolic Computer Use: Learning Reusable Policies for Reliable and Efficient Execution}},
  author = {Suh, Hyewon and Nguyen, Thanh Minh and Lee, Chih-Lun and Hartman, Darrow and Liu, Lizhao and Wang, Xin Eric and Li, Ang and Yang, Jiachen},
  year = {2026},
  eprint = {2609.36927},
  archivePrefix = {arXiv},
  doi = {10.48550/arXiv.2609.36927},
  url = {https://arxiv.org/abs/2609.36927}
}

@misc{verifier_hardening,
  title = {{Scalable Weak-to-Strong Verifier Hardening for Agent Environments}},
  author = {Zeng, Chong and Zhong, Ziqian and Zhang, Kexun and Bercovich, Ivan and Segal, Ivgeni and Mao, Yuzhen and Saxena, Shashwat and Raghunathan, Aditi},
  year = {2026},
  eprint = {2610.15924},
  archivePrefix = {arXiv},
  doi = {10.48550/arXiv.2610.15924},
  url = {https://arxiv.org/abs/2610.15924}
}

@misc{verifierbench,
  title = {{VerifierBench: Benchmarking the Generation of Executable Verifiers for LLM Agents}},
  author = {Yin, Jianing and Chen, Ziyi and Wang, Hao and Sun, Yiyou and Jin, Zijian and Song, Dawn and Zhang, Daoan},
  year = {2026},
  eprint = {2610.28679},
  archivePrefix = {arXiv},
  doi = {10.48550/arXiv.2610.28679},
  url = {https://arxiv.org/abs/2610.28679}
}

@misc{clift,
  title = {{CLIFT: Conformal Self-Verification for Web Agent Training and Test-Time Scaling}},
  author = {Zhang, Yifan and Dai, Yutong and Prabhu, Viraj and Hu, Zhiyuan and Xu, Ran and Chen, Zeyuan},
  year = {2026},
  eprint = {2610.06829},
  archivePrefix = {arXiv},
  doi = {10.48550/arXiv.2610.06829},
  url = {https://arxiv.org/abs/2610.06829}
}

@misc{leni_reliability,
  title = {{Where Does Agent Reliability Come From? A Cross-Benchmark Decomposition of Verification Loops, Specialist Models, and Scaffolding in a Production Enterprise Agent}},
  author = {Dastidar, Arunabh},
  year = {2026},
  eprint = {2607.17044},
  archivePrefix = {arXiv},
  doi = {10.48550/arXiv.2607.17044},
  url = {https://arxiv.org/abs/2607.17044}
}

@misc{sauron,
  title = {{Evidence-First Agentic Verification}},
  author = {Sanayei, Reza and Dumitru, Razvan-Gabriel and Hou, Sunjie and Rezaei, MohammadHossein and Riaz, Haris and Vacareanu, Robert and Yang, Minglai and Liu, Ying and Surdeanu, Mihai and Zhang, Daniel Yue},
  year = {2026},
  eprint = {2610.17724},
  archivePrefix = {arXiv},
  doi = {10.48550/arXiv.2610.17724},
  url = {https://arxiv.org/abs/2610.17724}
}

@misc{llm_verifier,
  title = {{LLM-as-a-Verifier: A General-Purpose Verification Framework}},
  author = {Kwok, Jacky and Li, Shulu and Atreya, Pranav and Liu, Yuejiang and Jiang, Yixing and Finn, Chelsea and Pavone, Marco and Stoica, Ion and Mirhoseini, Azalia},
  year = {2026},
  eprint = {2607.05391},
  archivePrefix = {arXiv},
  doi = {10.48550/arXiv.2607.05391},
  url = {https://arxiv.org/abs/2607.05391}
}
